# HYBRID MOBILITY PREDICTION OF 802.11 INFRASTRUCTURE NODES BY LOCATION TRACKING AND DATA MINING

Biju Issac[1], Khairuddin Ab Hamid[2], C.E. Tan[3]

[1,2] Faculty of Engineering
University Malaysia Sarawak
94300 Kota Samarahan, Sarawak, Malaysia
[1] bissac@swinburne.edu.my, [2] khair@cans.unimas.my

[3] Faculty of CS and IT
University Malaysia Sarawak
94300 Kota Samarahan, Sarawak, Malaysia
cetan@fit.unimas.my

*Abstract* - **In an IEEE 802.11 Infrastructure network, as the mobile node is moving from one access point to another, the resource allocation and smooth hand off may be a problem. If some reliable prediction is done on mobile node's next move, then resources can be allocated optimally as the mobile node moves around. This would increase the performance throughput of wireless network. We plan to investigate on a hybrid mobility prediction scheme that uses location tracking and data mining to predict the future path of the mobile node. We also propose a secure version of the same scheme. Through simulation and analysis, we present the prediction accuracy of our proposal.**

## 1. INTRODUCTION

The mobility management in wireless networks covers the options for storing and updating the location information of mobile users who are connected to the system. An interesting topic of research in mobility management field is mobility prediction. Mobility prediction can be explained as the prediction of a mobile user's next movement where the mobile user is traveling between the nodes or access points of a wireless network. The predicted path or movement can help increase the efficiency of wireless network, by effectively allocating resources to the most probable access point (that can be the next point of attachment) instead of blindly allocating excessive resources in the mobile nodes neighborhood of a mobile user (Saygin and Ulusoy, 2002), (Gok and Ulusoy, 2000). This paper is organized as follows. Section 2 gives the details of existing mobility prediction schemes, along with location tracking and data mining. Section 3 shows the mobility prediction proposal, section 4 is a secure version of that proposal, section 5 is simulation results and section 6 is the conclusion.

## 2. RELATED CONCEPTS AND WORK

A mobility model should be made to capture the movements of real mobile nodes (MNs). Changes in speed and direction of MN should happen in reasonable time slots. We discuss seven different mobility models for wireless networks (that are ad-hoc network oriented), which can as well be applied to infrastructure networks:

- Random Walk Mobility: A simple mobility model that is based on random directions and speeds.
- Random Waypoint Mobility: A model that includes pause times between changes in destination and speed.
- Random Direction Mobility: A model that forces MNs to travel to the edge of the simulation area before changing direction and speed.
- A Boundless Simulation Area Mobility: A model that converts a two dimensional rectangular simulation area into a torus-shaped simulation area.
- Gauss-Markov Mobility: A model that uses one tuning parameter or variable to vary the degree of randomness in the mobility pattern.

- Probabilistic Random Walk Mobility: A model that utilizes a set of probabilities to determine the next position of an MN.
- City Section Mobility: A simulation area that represents streets within a city.

*2.1 Random Walk*

Many entities in nature move in extremely unpredictable ways, and Random Walk Mobility was developed to mimic this erratic movement (Davies, 2000). By choosing a random direction and traveling speed, an MN moves from its current position to a new position. The new speed and direction are both chosen from pre-defined ranges. If a mobile node reaches a simulation boundary, it "bounces" off the simulation edge or border with an angle that is dependent on the incoming direction. The MN then continues along this new path. The Random Walk Mobility Model is a memory-less mobility pattern because it retains no knowledge concerning its past locations and speed (Liang and Haas, 1999).

*2.2 Random Waypoint*

The Random Waypoint Mobility Model includes pause times between changes in direction and/or speed (Johnson and Maltz, 1996). A mobile node begins by pausing in one location for a certain period of time. Once this pause time finishes, the MN chooses a random destination in the simulation area and a speed that is uniformly distributed between a minimum and maximum value, [min_speed, max_speed]. The mobile node then travels toward the new random destination at the speed that was selected. Upon its arrival, the mobile node pauses for a specified time period before repeating the cycle. We note that the movement pattern of an MN using the Random Waypoint Mobility Model is similar to the Random Walk Mobility Model if pause time is zero and [min_speed, max_speed] = [speed$_{min}$, speed$_{max}$]. In most of the performance analysis that use the Random Waypoint Mobility Model, the mobile nodes are distributed randomly. This initial random distribution of MNs has nothing to do with the manner in which nodes distribute themselves when moving.

*2.3 Random Direction*

To tackle the density waves in the average number of neighbours produced by the Random Waypoint Mobility Model, the Random Direction Mobility Model was created. A density wave is the clustering or grouping of nodes in a specific part of the simulation area. In the case of the Random Waypoint Model, this clustering occurs near the center of the simulation area. In the Random Way Point Model, the probability of a mobile node choosing a new destination that is located in the center of the simulation area is quite high. Thus, the mobile node path appears to converge and disperse again and again. In order to remove this strange behavior and to encourage a semi-constant number of neighbors throughout the simulation, the Random Direction Mobility Model was developed (Royer et al., 2001). In this model, mobile nodes choose a random direction in which to travel similar to the Random Walk Mobility Model. A mobile node then travels to the border of the simulation area in that direction. Once the simulation boundary is reached, the MN pauses for a specified time, chooses another angular direction (between 0 and 2π degrees) and continues the movement.

*2.4 A Boundless Simulation Area*

In the Boundless Simulation Area Mobility Model, there is a relationship between the previous direction of travel and velocity of a mobile node with its current direction of travel and velocity (Hass, 1997). A velocity vector v = (v, θ) is used to describe an MN's velocity v as well as its direction θ; The MN's position is represented as (x, y). Both the velocity vector and the position are updated at every Δt time steps according to the following formulas:

$$\begin{aligned} v(t+\Delta t) &= \min[\max(v(t)+\Delta v, 0), V_{max}]; \\ \theta(t+\Delta t) &= \theta(t) + \Delta \theta; \\ x(t+\Delta t) &= x(t) + v(t) * \cos\theta(t); \\ y(t+\Delta t) &= y(t) + v(t) * \sin\theta(t); \end{aligned} \qquad (1)$$

where *V$_{max}$* is the maximum velocity defined in the simulation, *Δv* is the velocity change that is uniformly distributed between *[−A$_{max}$ *Δt, A$_{max}$ *Δt]*, *A$_{max}$* is the maximum acceleration of a given MN, *Δθ* is the change in direction which is uniformly distributed between *[−α * Δt, α *Δt]*, and *α* is the maximum angular change in the direction a mobile node is traveling. The Boundless Simulation Area Mobility Model is also different in how the boundary of a simulation area is dealt with. In all the mobility models discussed, MNs reflect off or stop moving once they reach a simulation boundary or edge. In Boundless Simulation Area Mobility Model, MNs that reach one side of the simulation area continue traveling and reappear on the opposite side of the simulation area. Thus MN can travel unhindered as this technique creates a torus-shaped simulation area.

*2.5 Gauss-Markov*

The Gauss-Markov Mobility Model was proposed to handle different levels of randomness through one tuning parameter. Initially each MN is assigned a current speed and direction. At fixed intervals of time – n, the speed and direction of each MN are changed and the mobile node moves. Specifically, the value of speed and direction at the $n^{th}$ instance is calculated based upon the value of speed and direction at the previous $(n-1)^{st}$ instance and a random variable using the following equations:

$$s_n = \alpha s_{n-1} + (1-\alpha)\bar{s} + \sqrt{(1-\alpha^2)} s x_{n-1}$$
$$d_n = \alpha d_{n-1} + (1-\alpha)\bar{d} + \sqrt{(1-\alpha^2)} d x_{n-1}$$
(2)

where $s_n$ and $d_n$ are the new speed and direction of the MN at time interval n; α, where $0 \leq \alpha \leq 1$, is the tuning parameter used to vary the randomness; $s$ and $d$ are constants representing the mean value of speed and direction as n →∞; and $s_{xn-1}$ and $d_{xn-1}$ are random variables from a Gaussian distribution. Totally random values (or Brownian motion) are obtained by setting $\alpha = 0$ and linear motion is obtained by setting α = 1 (Liang and Haas, 1999). Intermediate levels of randomness are obtained by varying the value of α between 0 and 1. At each time interval the next location is calculated based on the current location, speed, and direction of movement. Specifically, at time interval n, an MN's position is given by the equations:

$$x_n = x_{n-1} + s_{n-1} \cos d_{n-1}$$
$$y_n = y_{n-1} + s_{n-1} \sin d_{n-1}$$
(3)

where $(x_n, y_n)$ and $(x_{n-1}, y_{n-1})$ are the *x* and *y* coordinates of the MN's position at the *nth* and $(n-1)^{st}$ time intervals, respectively, and $s_{n-1}$ and $d_{n-1}$ are the speed and direction of the MN, respectively, at the $(n-1)^{st}$ time interval. To ensure that an MN does not remain near an edge of the grid for a long period of time, the MNs are forced away from an edge when they move within a certain distance of the edge. This is done by modifying the mean direction variable d in the above direction equation.

*2.6 Probabilistic Random Walk*

Chiang's mobility model uses a probability matrix to find the position of a particular MN in the next time step, which is denoted by three different states for position x and three different states for position y (Chiang, 1998). State 0, 1 and 2 are defined as follows: State 0 represents the current (x or y) position of a given MN, state 1 represents the MN's previous (x or y) position, and state 2 represents the MN's next position if the MN continues to move in the same direction. The probability matrix used is:

$$P = \begin{bmatrix} P(0,0) & P(0,1) & P(0,2) \\ P(1,0) & P(1,1) & P(1,2) \\ P(2,0) & P(2,1) & P(2,2) \end{bmatrix}$$

where each entry *P(c, d)* represents the probability that an mobile node will go from state c to state d. The values within this matrix are used for updates to both the MN's x and y position. In the simulator developed by Chiang, every node moves randomly with a predefined average speed. The following matrix contains the values Chiang used to calculate x and y movements:

$$P1 = \begin{bmatrix} 0 & 0.5 & 0.5 \\ 0.3 & 0.7 & 0 \\ 0.3 & 0 & 0.7 \end{bmatrix}$$

*2.7 City Section Mobility Model*

In the City Section Mobility Model, the simulation area imitates a street network that represents a portion of a city that has wireless network (Davies, 2000). The type of city being simulated dictates the streets and speed limits on the streets.. For example, the streets may form a grid in the downtown area of the city with a high-speed highway near the border of the simulation area to represent a loop around the city. Each MN begins the simulation at a defined point on some street. An MN then randomly chooses a destination, also represented by a point on some street. The movement algorithm from the current destination to the new destination locates a path corresponding to the shortest travel time between the two points; in addition, safe driving characteristics such as a speed limit and a minimum distance allowed between any two MNs exist. Upon reaching the destination, the MN pauses for a specified time and then randomly chooses another destination (i.e., a point on some street) and repeats the process. The City Section Mobility Model provides realistic movements for a section of a city since it severely restricts the traveling behavior of MNs. In other words, all MNs must follow predefined paths and behavior guidelines (e.g. traffic laws).

*2.8 Location Tracking and Data Mining*

Liu et al. (1998) have proposed a mobility modeling, location tracking, and trajectory prediction in Wireless ATM Networks. This paper treats the problem by developing a hierarchical user mobility model that closely represents the movement behavior of a mobile user, and that, when used with appropriate pattern matching and Kalman filtering techniques, yields an accurate location prediction algorithm, which provides necessary information for advance resource reservation (Liu et al., 1998). Capkun and collaborators (2001) propose a distributed, infrastructure-free positioning algorithm that does not rely on Global Positioning System (GPS). The algorithm uses the distances between the nodes to build a relative coordinate system in which the node positions are computed in two dimensions (Capkun et al., 2001). Yavas and colleagues (2005) propose a new algorithm for predicting the next inter-cell movement of a mobile user in a Personal Communication Systems network. In the first phase of the three phase algorithm, user mobility patterns are mined from the history of mobile user trajectories. In the second phase, mobility rules are extracted from these patterns, and in the last phase, mobility predictions are accomplished by using these rules (Yavas et al., 2005).

## 3. MOBILITY PREDICTION PROPOSAL

We propose a centralized mobility prediction technique that is a hybrid version of location tracking without GPS and data mining technique called *Location Tracking with Data mining Prediction Scheme* (LTDPS). Location tracking is done through a central server that receives the data regarding neighbouring APs from the mobile node. Data mining is done by the same server by using the mobility path history of the mobile node movements and by extracting the mobility patterns from that. We adapted some ideas from (Yavas et al., 2005), though implemented it differently.

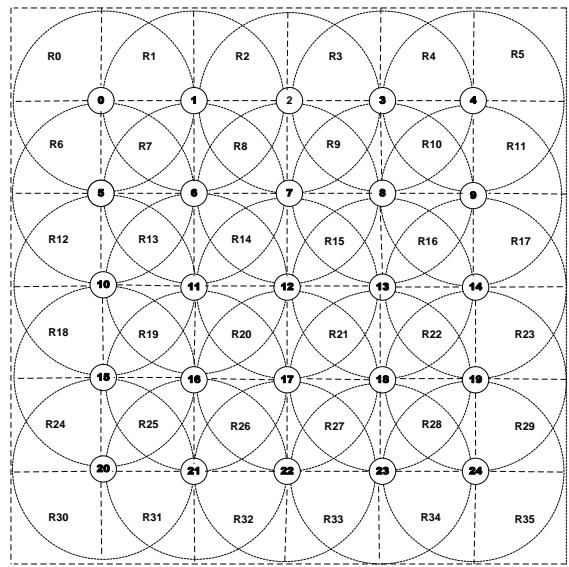

Figure 1. Access Points and Mobile regions arranged in a matrix format.

One of the significant assumptions made is that the mobile node once it starts moving in one direction will not abruptly change its direction. The mobility model used is a modified Random Waypoint model (Johnson and Maltz, 1996), where the probability of choosing a center AP in the 5 x 5 grid, is not that high. Consider the access points numbered 0 to 24, arranged in a 5 x 5 matrix form. Surrounding each access point (AP) there are four regions. These regions are numbered from R0 to R35 and is positioned around the 25 access points, as a 6 x 6 matrix as shown in the Figure 1. The mobile node can move around any AP in the surrounding four regions or can move to other adjacent regions of neighbouring AP.

A Mobile Path Prediction Server (MPPS) is connected to the Local Area Network where the Access Points are connected, to make the mobility prediction in a centralized way as shown in Figure 2.

At frequent intervals of time, the mobile node transmits the Received Signal Strength Indication (RSSI) of the surrounding APs to the MPPS. The IEEE 802.11 standard defines a way by which RF energy is to be measured by the circuitry on a wireless network interface card (NIC). This value is an integer in the range of 0-255 (a 1-byte value) called the Receive Signal Strength Indicator (RSSI). None of the vendors have measured the 256 different signal levels, and hence there is a standard discrepancy in that each vendor's 802.11 NIC will have a specific maximum RSSI value called, "RSSI_Max". For instance, Cisco chooses to measure 101 separate values for RF energy, and their RSSI_Max is 100. Other companies such as Symbol uses an RSSI_Max value of 31 and Atheros chipset uses an RSSI_Max value of 60 (Wild Packets, 2002). Essentially the communication by MN to MPPS server involves the AP's identity and the received signal strength. The MN associates at any time with the AP whose received signal strength is the highest.

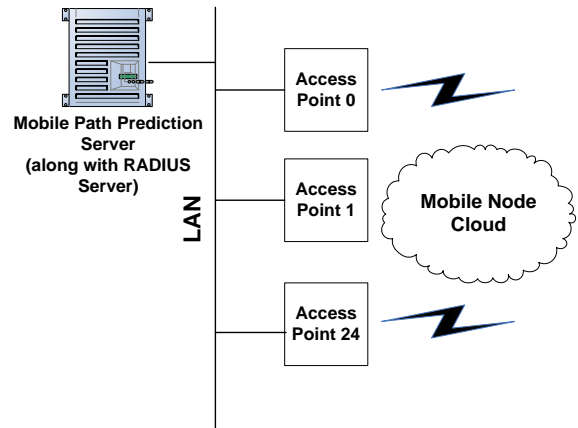

Figure 2. Centralized Server for Mobility Prediction

*3.1 Location Tracking*

We assume that the MN moves from one region to the other and pauses for a duration of t time units in any region, before making any movements further. By seeing the details of communication of AP's identity and RSSI value, the MPPS server is able to know which region the mobile node is located. All valid mobile paths by different users or mobile nodes are recorded (along with regions) in a central database in MPPS server.

Consider a section of the grid as shown in Figure 3. This is shown here for more clarity in our forthcoming discussions. For example, if the MN transmits details of AP numbered 0, 1, 5 and 6, MPPS server knows that the MN is located in region 7 (R7), according to Figure 1 and 3. Moreover, if the RSSI value (units) can be of the order as shown below: AP {0, 1, 5, 6} = RSSI {20, 30, 10, 5}, the MN is closer to AP1 and would be associated to AP1 in region 7 (R7). So the mobile node path is: 1. It also shows that MN is also closer to AP0, compared to AP5 and AP6. After the starting AP, only when the MN is mobile, it transmits RSSI information. We call this information as RSSI sample. The MN is able to know it is mobile when the RSSI sample that it receives from its neighbouring AP's differs widely. An error factor of $\Delta e$ can be allowed in RSSI values received by MN and transmitted to MPPS server. MN does the RSSI transmission to server in $\Delta t$ time units frequency interval. By analyzing the RSSI sample, the MPPS server is able to know the direction of MN movement. The region-AP map that the MPPS server stores can be shown in Table 1 for the first 12 regions, based on Figure 1.

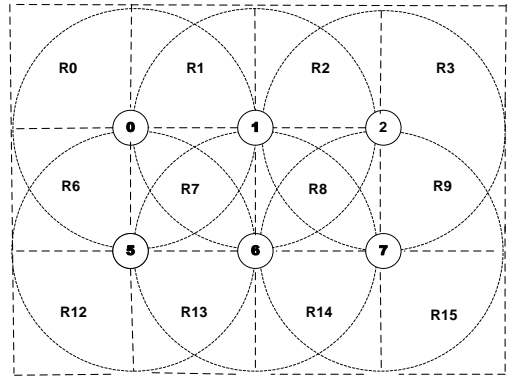

Figure 3. An enlarged view of a section of the grid shown in Figure 1.

For example, if MN moves from region 7 and the RSSI sample transmitted can contain – RSSI {5, 38, 15, 25} corresponding to AP {0, 1, 5, 6}. This shows that MN is closer to AP1 and AP6 and is moving toward region 8. When it enters the region 8, the RSSI sample can be: RSSI {5, 10, 20, 35} corresponding to AP {1, 2, 6, 7}. When the RSSI sample received by MPPS server is quite similar in value or when current AP signal is weak, the MPPS server would send signals to reserve resources (as explained later) in AP7 (with highest RSSI value) and would direct MN to initiate the handoff to AP7. If the handoff is successful, the mobile path is 1→7. Here the prediction would be accurate and the handover is initiated by the MPPS server.

*Location Tracking Algorithm:*

1. Initially MN is connected to an AP to start with, which would be located in Region R$i$, where i=0 to 35. MN would be tracking RSSI samples of all its neighboring AP's during $\Delta t$ time intervals.

2. The MN sends the RSSI samples (four highest values) to MPPS server, after its first connection is settled. MPPS server now knows the region where MN is located. The MN or the server knows that MN is stationary when the RSSI samples that it gets remain quite constant.

Table 1. The region-AP map for location tracking

| No. | Region | Neighbour APs |
|---|---|---|
| 1. | 0 (corner) | {0} |
| 2. | 1 | {0, 1} |
| 3. | 2 | {1, 2} |
| 4. | 3 | {2, 3} |
| 5. | 4 | {3, 4} |
| 6. | 5 (corner) | {5} |
| 7. | 6 | {0,5} |
| 8. | 7 | {0, 1, 5, 6} |
| 9. | 8 | {1, 2, 6, 7} |
| 10 | 9 | {2, 3, 7, 8} |
| 11 | 10 | {3, 4, 8, 9} |
| 12 | 11 | {4, 9} |

3. When there is difference in RSSI samples received, the MN or server knows that MN is on the move and sends that changed RSSI samples to MPPS server. The MN movement can be within the same region, and it may get connected to a closer AP whose RSSI value is highest, where the received RSSI samples would be {LV, LV, LV, HV} in any order. MPPS server checks the direction of MN movement, by analyzing the RSSI sample. At any time, if the RSSI value is as in Table 2, the MPPS server could sense MN's direction of movement and the process of resource reservation can be initiated by MPPS server.

Table 2. Location Tracking through RSSI samples

| No. | RSSI Sample | Direction/Position of Mobile Node |
|---|---|---|
| 1. | {MV, MV, MV, MV} | Middle of any Region |
| 2. | {LV, LV, HV, HV} | Movement toward the center of APs with HV values and horizontal or vertical cross over results. |
| 3. | {LV, LV, LV, HV} | Confirmed movement toward the AP with HV value. |

4. MPPS server knows what would be the next region that MN would enter, based on RSSI sample value – whether high value (HV), medium value (MV) or low value (LV) as in Table 2. It immediately has information to predict where the MN's next attachment AP would be, based on the following formula. If the MN moves from one region (current region) to the other region (next region), apply the formula: *(Current AP {all neigbour APs}) ∩ Predicted Region {all APs} = {S}*, where S is the set of probable next APs. If S contains only one element, then that is the next predicted AP or if it contains 2 or more elements, MPPS server does the prediction based on data mining to get the best possible path. Data mining concept is explained in next section.

As shown in Figure 4, the MN can move vertically, diagonally or horizontally (roughly, even though it depends on the source point of traversal). This movement can be tracked by MPPS server. The RSSI values which can be classified as low value (LV), medium value (MV) and high value (HV) is mapped on a 3 x 3 matrix region within region 8, for MN connecting to AP6 from AP5, as in Figure 4.

We understand that there could be RSSI noise, which may need filtering, such as Kalman, Gray or Fourier filtering. But for the moment we are not considering the filtering options. There can be some indecisive scenarios. They are as follows – the RSSI values received are very weak or corrupted from all neighbouring APs and no clear decision can be taken on the next AP, the RSSI values received are similar in value, after accounting in the error factor of ∆e and the MN can still move. The RSSI values are classified equal or not by checking whether it falls within a range of values (such as, 30-35 units) and not necessarily exact integer or decimal values.

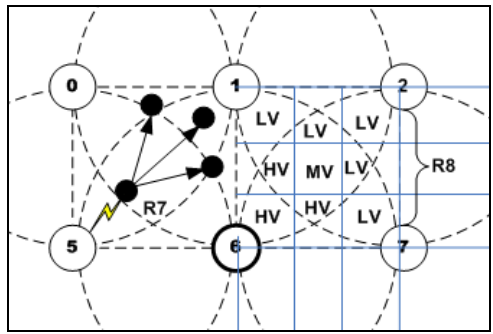

Figure 4. Mobile Node (Dark circle) connected to AP5 in region 7 moves (vertically, diagonally or horizontally) and LV, MV and HV RSSI values in region 8.

*3.2 Data Mining*

To counter the above indecisive scenarios, we incorporate the data mining approach in parallel, once MPPS server knows the MN location, through RSSI samples sent. The Data mining approach consists of three phases: user mobility pattern mining, generation of mobility rules using the mined user mobility patterns, and the mobility prediction. The next inter-cell movement of mobile users is predicted based on the mobility rules in the last phase. For example, if the MN is in region 8 (R8), and connected to AP1, the only possible movements are toward AP {2, 6, 7}. Using the extracted mobility patterns (for node movements: 1→2, 1→6, 1→7, 1→2→7, 1→6→7) from the database of mobile paths in MPPS server, next AP prediction is done. If the RSSI sample received has similar values such as –RSSI {10, 12, 30, 30} for AP {1, 2, 6, 7}, then the MPPS server goes in for data mining approach, as MN can still move. The values need not be 30 as listed for AP6 and AP7. It can be 30 and 35 and with an error factor of 5, the values are counted similar.

*3.2.1 User Mobility Pattern mining*

Once the MPPS server knows the location of MN, it locates all the MN's immediate neighbours. Consider for example a path: 19(22)→13(15)→8(9)→9(10)→4(4) →3(3), where values in parenthesis represent the regions where MN is present. The starting AP is AP19 and the region is 22. The MN moves to region 15 then and gets associated to AP13. Once the MN moves from one region (current region) to the other region (next region), we need to apply the formula: *Current AP {all neigbhour APs} ∩ Next Region {all APs} = {S}*, where S is the set of probable next APs. To find the predicted next AP, as per the rule given above, take the intersection of neighbour AP set of AP19 and neighbour AP set of region 15. This is given as: AP19 {neigbours} ∩ R15 = {13, 14, 18, 23, 24} ∩ {7, 8, 12, 13} ={13}. Thus the predicted next AP is 13, which is vaild as the MN movement. Predicted path = 19→13. From AP13 in region 15, the MN moves to region 9 and gets associated to AP8. AP13 {neigbours} ∩ R9 ={7, 8, 12, 9, 14, 17, 18, 19}∩ {2, 3, 7, 8} = {7, 8}. Since we have 2 entries 7 and 8, we need to do pattern mining. From AP13, the MN can move to AP7 or AP8. So 13→7 and 13→8 patterns are searched in the sorted database and their frequency count is noted as in Table 3. The path 13→8 is selected because of higher frequency count. AP8 is the next predicted AP. Predicted path = 13→7→8. This procedure is repeated for other nodes.

   Consider a path 22(26)→21(25)→15(18)→16(19). Following our earlier calculation, AP22{neigbours} ∩ R25 ={16, 17, 18, 21, 23} ∩ {15, 16, 20, 21} ={20, 21}. Here MN moves from AP22 in region 26 to AP21 in region 25, as 22→21 has higher frequency count (of 272) than 22→16 (where the count was 162), like we discussed before. Predicted path is vaild and thus is: 22→21.

   Then it moves to region 18. So, AP21{neigbours} ∩ R28 ={15, 16, 17, 20, 22} ∩ {10, 15} ={15}. So the MN moves to AP15 in region 18. Predicted path is vaild and thus is: 22→21→15. From there it moves to region 19. Following our earlier calculation again, AP15{neigbours} ∩ R19 ={10, 11, 16, 20, 21} ∩ {10, 11, 15, 16} ={10, 11, 16}.

Table 3. Mobility Pattern mining for 2 neighbours

| No. | Path | Frequency |
|---|---|---|
| 1 | 13→7 | 238 |
| 2 | 13→8 | 367 |

   Note there are there possible movements here, which are 15→10, 15→11 and 15→16. In such a situation, we would like to do the data mining, slightly differently [12] as in Table 4. The final score for 15→10 = 707 + x1*0.5 + x2*0.5, for 15→11 = 40 + x3*0.5 + x4*0.5 and for 15→16 = 54 + x5*0.5 + x6*0.5. Here, we are taking a corruption factor of 0.5 that has to be multiplied with the 'indirect path counts'. When we consider the path 15→10, paths such as 15→11→10 and 15→16→10 are termed as indirect paths, as they have the source and destination APs as vaild. Total score = score of 2 AP link + corruption factor * {sum of scores of 3 AP link}. In this case, we found our prediction not vaild (to be from 15→10, instead of the vaild path 15→16). But the concept will definitely work for other cases as the mobile path database gets populated.

Table 4. Mobility Pattern mining for 3 neigbours

| No. | Path | Frequency |
|---|---|---|
| 1 | 15→10 | 707 |
| 2 | 15→11→10 | x1 |
| 3 | 15→16→10 | x2 |
| 4 | 15→11 | 40 |
| 5 | 15→10→11 | x3 |
| 6 | 15→16→11 | x4 |
| 7 | 15→16 | 54 |
| 8 | 15→10→16 | x5 |
| 9 | 15→11→16 | x6 |

*3.2.2 Generation of Mobility Rules*

An example of mobility rule generation from our previous examples is given in Table 5. Mobility rule can thus be generated as in Table 5, after the mining is done for specific paths. These mobility rules differ depending on the direction of MN movement. Since we incorporate the concept of region, once a MN enters a region, it gets associated to one of the neighbouring APs, depending on the direction of node movement. That narrows down the probability of error as the movement can happen to 1, 2 or 3 APs, depending on which region MN is located.

Table 5. Mobility rule generation

| No. | Region Movement | Predicted Path |
|-----|-----------------|----------------|
| 1   | AP13:R15→R9     | AP8 (valid)    |
| 2   | AP22:R26→R25    | AP21 (valid)   |
| 3   | AP15:R18→R19    | AP10 (invalid) |

*3.2.3 Prediction of Mobility*

Once the Mobility rules are generated, the predicted next AP is the one whose path has the highest frequency. In the example that we are discussing, based on Table 5, the next predicted AP would be AP8, as per the first rule in Table 5. The MPPS server would send signals to reserve resources in AP8 and would allow MN to initiate the handoff to AP8. To optimize the performance, in case of invalid prediction, the two highest frequencies can be noted and resources can be reserved on two APs. We haven't considered that option in our simulation.

*3.3 Overall Mobility Prediction Algorithm*

The overall mobility prediction algorithm is stated below that includes location tracking and data mining.
1. Initially the Mobile Node (MN) starts by associating with a random Access Point (AP).

2. The MN transmits the RSSI value received and the surrounding AP's identity (maximum of 4) to MPPS Server. MN stays with that AP for t time units, where t is variable. To be precise, MN sends a packet with <RSSI_NextAP, NextAP identity, RSSI_CurrentAP, CurrentAP identity, MN identity> value to Server for all the surrounding APs, where RSSI_NextAP and RSSI_CurrentAP is the RSSI values of neighbouring AP and currently attached AP to MN respectively. As each region can have maximum 4 APs and minimum 2 (or 1) APs, the right AP identity is ensured by checking the RSSI value. This transmission happens in regular intervals; for example, Δt seconds/time units.

3. For a given MN identity, the server would choose four packets with the highest RSSI values (for Next AP) and note the AP identity (even if the RSSI values match). In the four RSSI values selected, if all the values are above the Next AP RSSI threshold value ("region threshold" RSSI value, $RSSI_{region\_threshold}$), then the MN is in the inner region with four neighbouring APs, or if only two values are above the Next AP RSSI threshold value, then the MN is in the outer region with two neighbouring APs, or if only one value is above the Next AP RSSI threshold value, then the MN is in the corner region with only one neighbouring AP. Thus, analyzing the Next AP RSSI values by the Server would indicate to which AP the MN is getting closer. If one of the RSSI value is highest then the MN is closer to the corresponding AP, or if two of the highest RSSI values (or block values) are the same, then the MN is equally closer to those two APs, or if three or four of the RSSI values are the same, then the MN is in the center of a given region.

4. After t time units, the MN starts moving and it transmits RSSI sample to MPPS Server. Now the direction of movement of MN can be known to MPPS server, through the RSSI sample values that it receives after Δt time units, as explained before. If "n" packets sent by MN are same (having RSSI values same or closer values), then the server knows that MN is stationary or moving closer to a specific AP for a period of time.

5. The MPPS server adopts data mining (that happens in parallel) to predict the next AP of MN attachment, when the RSSI sample values received in a set are similar (such as, for two or more APs) or when the RSSI samples is weak or no signals are coming from neigbouring APs. Here it uses the mobile node's past history to extract mobility patterns, whereby it can come to a good judgment, as explained in section 3.2. Once the MN moves from one region (current region) to the other region (next region), we need to apply the formula: *Current AP {all neighbor APs} ∩ Next Region {all APs} = {S},* where S is the set of probable next APs. If S contains only one element, then that is the next predicted AP or if it contains 2 or more elements, it adopts data mining to get the best possible path.

6. The MPPS server intimates the selected AP (selected with high RSSI value or selected through data mining) at an optimum hand-off time ($t_{opt\_handoff}$), that a particular MN is trying to attach to it and informs about the resources needed to be reserved, which is the first-stage reservation. A copy of this notice is sent to MN also, optionally. In the next Δt time units, if the MN senses that the Next AP RSSI value is equal to or greater than the threshold RSSI value ($RSSI_{region\_threshold}$), the MN sends a second-stage reservation to the AP, according to the type of traffic like data, voice and video. The voice and video would imply more buffer space in the second stage reservation request, unlike data that needs fewer buffers.

7. Hand off can be initiated through RSSI sample comparison or data mining. If the MN senses that the Next AP RSSI value is equal to or greater than the region threshold RSSI value ($RSSI_{region\_threshold}$) and at the same time, the current AP RSSI value is diminishing, hand off can be initiated from Current AP to Next AP through MPPS server. During the hand-off, the current AP's permission is sought to allow MN to get connected to Next AP.

8. Once the handoff is successful, the mobile node path is stored in the MPPS database.

Note: If the MN at any point is sensing that it is losing the network connection from the current AP, because of some delay in the communication from MPPS server, the MN can initiate the hand off to the new AP. This is can be a backup option.

# 4. SECURE MOBILITY PREDICTION

We also propose a secure version of the above algorithm using a shared secret key, as the process is prone to message modification, spoofing and replay attacks by rouge mobile nodes. Message modification and spoofing attack happens when the attacker modifies the content (RSSI value or others) and identity of the AP or MN respectively. Replay attack occurs when the attacker replays the captured packet at a later time for defeating the prediction or for masquerading. The secure mobility prediction packet exchange can be as follows:

*For RSSI Communication initiated by MN:*
1. The packet send by MN is appended by a message integrity control (MIC). The MN encrypts the packet/message using 3DES with cipher block chaining (CBC). It cuts the message into predetermined-sized of i blocks (where i = 1, 2, … n). It then uses the CBC residue (that is the last block output by CBC process) as the MIC (as $MIC_1$, $MIC_2$ etc.), which would act as the checksum. The clear text message plus the MIC would be transmitted to the server.

2. The server encrypts the received plaintext message from MN using 3DES with the shared secret key ($K_{SA}$) and performs the hashing process to produce a similar MIC (such as, MIC*). The server then checks if the MIC (received from MN) = MIC*, and if that's true, then the message is non-tampered in transit. Otherwise, it is rejected.

3. Finally, the server sends an acknowledgement (ACK) and a random number (RND) or a nonce (as $RND_1$, $RND_2$ etc.) that is encrypted with the shared secret key $K_{SA}$, i.e. $(ACK+RND)K_{SA}$ to MN.

5. The MN verifies the encrypted ACK from server and the next message to MPPS server (after Δt time units) is similar to the initial communication, but with the addition of encrypted RND or nonce. If the server or MN at any time receives any unauthorized message, an alarm can be generated or logged and connection refused for that attempt. When the server responds back to MN with an $(ACK+RND)K_{SA,}$ it would be with a different RND or nonce.

*For Communication initiated by the MPPS server:*
6. If the MPPS server needs to communicate a message to MN in relation to resource reservation or handoff, the same approach can be done in the reverse order.

The packet exchange described above (initiated by MN) that can stop the security attacks is shown below for the two sets of interactions, as a set of equations:

1. MN → MPPS          : $Message_1 + MIC_1$
2. MPPS → MN          : $(ACK+RND_1)K_{SA}$
3. MN → MPPS          : $Message_2 + (RND_1)K_{SA} + MIC_2$
4. MPPS → MN          : $(ACK+RND_2) K_{SA}$

If MPPS server needs to send a message to MN (i.e. initiated by server), it can be as follows:

1. MPPS → MN : $Message + (RND)K_{SA} + MIC$
2. MN → MPPS  : $(ACK+RND)K_{SA}$

As we can observe, the security process involve only light encryption for authentication purposes and an additional acknowledgement as overhead. The actual messages are sent as plain text to reduce processing overhead.

## 5. SIMULATION RESULTS

The simulation is done with the following parameters. A modified version of Random Waypoint Mobility model is used, where nodes moves equally to the sides and center. The mobile node travels in a 6 x 6 region matrix which contains 5 x 5 access point matrix. Thus the 36 regions have 25 access points placed in a matrix format as shown in Figure 1.

The simulation is run to create 10,000 mobile paths and this is stored in a central database of MPPS server. This data set is used for data mining and to create the mobility patterns. Later, a test set of 10 random mobile paths were created (with a maximum of six access points and a minimum of three access points) and tested for prediction accuracy. The reasoning between the number of paths and the number of access points within them is that there would be a minimum of thirty mobile node transitions from one access point to another (36 transitions, as per Table 6) that would happen, which would make our calculations statistically significant. The accuracy of prediction through RSSI sample approach and location tracking is 100%, if the RSSI value received by MN is high from a particular AP, during its mobile path. The data mining approach which was done in parallel was simulated and the prediction analysis is as given below.

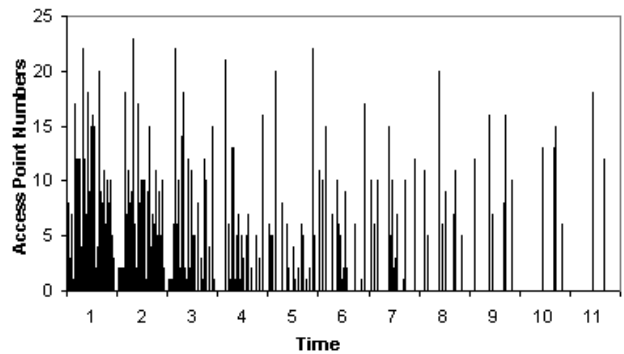

Figure 5. Mobile Node Paths generated (only a sample of around 30 paths is shown).

We would like to show the prediction accuracy for a sample of 10 paths, as given below. The mobile node movement history consisted of 10,000 mobile paths.

Path 1:  7(8) →2(2) → 6(7) →1(1)
Path 2:  13(15) → 7(8) → 1(1) → 0(0) → 6(7)
Path 3:  23(27) → 17(20) →16(19) →11(13)→ 6(7)
Path 4:  22(26) → 21(25) → 15(18) → 16(19)
Path 5:  11(13) → 5(6) → 0(0)
Path 6:  14(16) → 9(10) → 8(9) → 2(2) →1(1) → 0(0)

Path 7:   17(20) → 12(14) → 6(7) →11(13)
Path 8:   12(14) → 11(13) → 5(6) → 6(7)
Path 9:   8(9) → 7(8) → 2(2) → 1(1) → 0(0)
Path 10: 19(22) → 13(15) → 8(9) → 9(10) →4(4)→3(3)

The values in the parenthesis are the region numbers. i.e. 7(8) means AP7 in region 8. As a result of the simulation, the Table 6, 7 and 8 shows the prediction accuracy for the random sample considered above, using three different prediction schemes. The average accuracy of LTDPS was 76.4%. The predicted path contains, for example entries such as 0(1, 2) in row 1 of Table 6. The 0 is the predicted next AP (AP0) and inside the parenthesis, 1 is the actual node to which MN moved and 2 is the frequency rank of the predicted node, which is AP0. The variance between 33% (smallest value) and 100% (highest value) in Table 6 is the result of invalid prediction in the paths.

Table 6. Predicted Path Accuracy Table for a sample using our prediction scheme (LTDPS)

| No  | Original Path | Predicted Path | Accuracy (%) |
|-----|---------------|----------------|--------------|
| 1.  | 7→2→6→1 | 7→1(2, 2)→6→0(1, 2) | 1 (33%) |
| 2.  | 13→7→1→0→6 | 13→7→1→0→5(6, 2) | 3 (75%) |
| 3.  | 23→17→16→11→6 | 23→17→16→11→ 5(6, 2) | 3 (75%) |
| 4.  | 22→21→15→16 | 22→21→15→10(16, 2) | 2 (67%) |
| 5.  | 11→5→0 | 11→5→0 | 2 (100%) |
| 6.  | 14→9→8→2→1→0 | 14→9→8→2→1→0 | 5 (100%) |
| 7.  | 17→12→6→11 | 17→11(12, 2)→6→11 | 2 (67%) |
| 8.  | 12→11→5→6 | 12→11→5→ 0(6, 2) | 2 (67%) |
| 9.  | 8→7→2→1→0 | 8→7→2→1→0 | 4 (100%) |
| 10. | 19→13→8→9→4→3 | 19→13→8→3(9, 2)→4→3 | 5 (80%) |

We compared our prediction method with two other methods, as baseline schemes. Refer to Tables 7 and 8. The first prediction method is called Mobility Prediction based on Transition Matrix (TM). In this method, a cell-to-cell transition matrix is formed by considering the previous inter-cell movements of the mobile nodes or users. The predictions are based on this transition matrix by selecting the *x* most likely cells or regions as the predicted cells. We used TM for performance comparison because it makes predictions based on the previous movements of the mobile node or user (Rajagopal et al., 2002). Assuming x=1 (as the previous scheme also used x=1), the average accuracy of TM was found to be 38.8% in our simulation.

The second prediction method is the Ignorant Prediction (IP) scheme (Bhattacharya and Das, 2002). This approach disregards the information available from movement history. To predict the next inter-cell movement of a user, this method assigns equal transition probabilities to the neighboring cells of the mobile nodes currently residence cell. It means that prediction is performed by randomly selecting *m* neighboring cells of the current cell. We have taken m to be the maximum no. of neighbours possible. The value in the parenthesis in the paths shows the corrected AP number. The average accuracy of this scheme was found to be 36.3% and as expected was quite inconsistent.

Table 7.  Predicted Path Accuracy Table for a sample using Transition Matrix (TM) Prediction

| No | Original Path | Predicted Path | Accuracy (%) |
|---|---|---|---|
| 1. | 7→2→6→1 | 7→2→1(6,4)→5(1,2) | 1 (33%) |
| 2. | 13→7→1→0→6 | 13→8(7,3)→2(1,2)→0→5(6,2) | 1 (25%) |
| 3. | 23→17→16→11→6 | 23→22(17,3)→16→15(11,2)→10(6,3) | 1 (25%) |
| 4. | 22→21→15→16 | 22→21→20(15,2)→10(16,3) | 1 (33%) |
| 5. | 11→5→0 | 11→10(5,2)→0 | 1 (50%) |
| 6. | 14→9→8→2→1→0 | 14→9→14(8,3)→3(2,2)→1→0 | 3 (60%) |
| 7. | 17→12→6→11 | 17→16(12,3)→11(6,3)→5(1,3) | 0 (0%) |
| 8. | 12→11→5→6 | 12→11→5→ 0(6,5) | 2 (67%) |
| 9. | 8→7→2→1→0 | 8→3(7,3)→2→1→0 | 3 (75%) |
| 10. | 19→13→8→9→4→3 | 19→14(13,8)→8→3(9,5)→ 14(4,2)→9(3,2) | 1 (20%) |

Table 8.  Predicted Path Accuracy Table for a sample using Ignorant Prediction

| No | Original Path | Predicted Path | Accuracy (%) |
|---|---|---|---|
| 1. | 7→2→6→1 | 7→1 (2)→1(6)→0(1) | 0 (0%) |
| 2. | 13→7→1→0→6 | 13→7→1→0→1(6) | 3 (75%) |
| 3. | 23→17→16→11→6 | 23→17→11(16)→10(11)→5(6) | 1 (25%) |
| 4. | 22→21→15→16 | 22→21→15→20(16) | 2 (75%) |
| 5. | 11→5→0 | 11→10(5)→0 | 1 (50%) |
| 6. | 14→9→8→2→1→0 | 14→8(9)→3(8)→3(2)→1→0 | 2 (40%) |
| 7. | 17→12→6→11 | 17→13(12)→7(6)→0(11) | 0 (0%) |
| 8. | 12→11→5→6 | 12→7(11)→5→0(6) | 1 (33%) |
| 9. | 8→7→2→1→0 | 8→2(7)→1(2)→7(1)→0 | 1 (25%) |
| 10. | 19→13→8→9→4→3 | 19→13→7(8)→2(9)→3(4)→3 | 2 (40%) |

A comparison bar graph that shows the prediction accuracy can be as shown in Figure 6, for the three different schemes.  It is very clear that LTDPS (our proposal) is having better accuracy and generally the accuracy of TM and IP tends to be lower.

The dividing of surrounding AP area into four regions plays a great role in making the proposal more accurate. This would reduce the number of possible next AP set and hence the prediction can be more accurate. If such a division is not made, then based on the network model used in Figure 1, there would be eight neighbours to an AP, and to predict one of them would be quite a difficult task. So the location tracking through RSSI samples, coupled with data mining makes our proposal work better. Note the higher bars that touch 100% a few times with LTDPS.

Another observation that can be made is the frequency rank of the nodes (shown in brackets) in the predicted path for LTDPS and TM from Table 6 and 7. From Table 6 for LTDPS, it is clear that the frequency rank is mostly 1 or 2. Frequency rank 1 means prediction is vaild (hence not shown in brackets in the Table). But from Table 7 for TM, the frequency ranks fluctuates between 1, 2, 3, 4 and 5, thus showing how it's lacking in accuracy. The maximum frequency deviation graphs for LTDPS and TM schemes can be as shown in Figure 7.

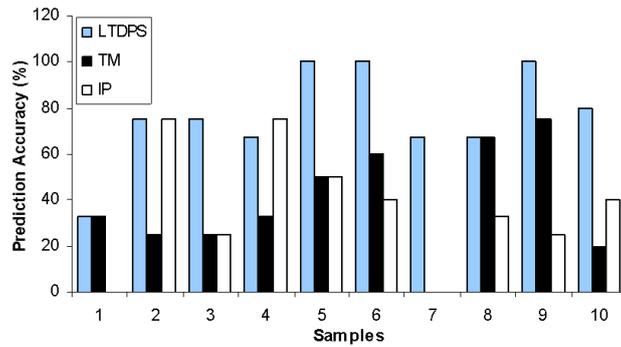

Figure 6. Prediction accuracy graph for 10 random sample paths for LTDPS, TM and IP Schemes

As per Figure 7, a curve or line closer to frequency rank 1 is better. TM does not touch the frequency rank of 1, as we are taking the maximum frequency deviation for all samples and in all paths at least one prediction is invalid. The line or curve would touch frequency rank 1 if and only if all the predicted paths are valid. For example, a frequency rank of 2 shows that, though it has missed the right AP prediction, it has chosen the AP that has the next highest frequency rank. Sample paths 5, 6 and 9 shows for LTDPS that all the paths predicted are valid.

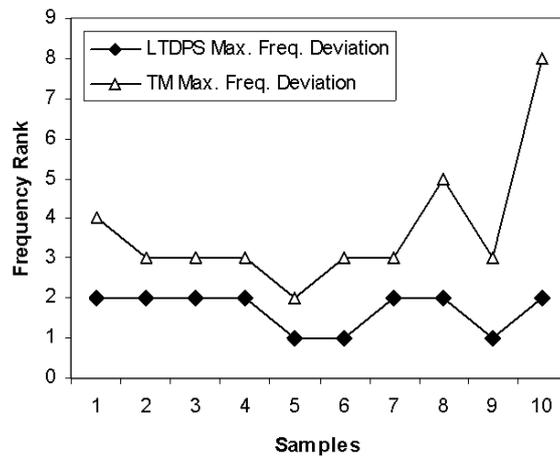

Figure 7. Max. Frequency Deviation graph for 10 random sample paths for LTDPS and TM Schemes. Frequency rank 1 shows accurate prediction.

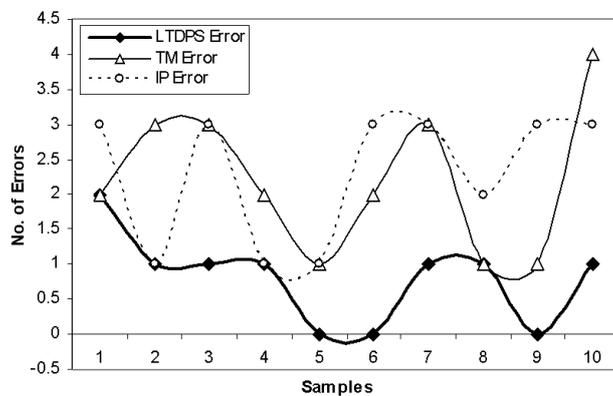

Figure 8. Error graph for 10 random sample paths for LTDPS, TM and IP Schemes.

Finally a curve plot of error margin of the three prediction schemes can be as shown in Figure 8. Here we don't consider the number of AP transitions by the MN, but only the path errors in prediction. The closer, the line or curve is to the x-axis, the better the scheme is.

## 6. CONCLUSION

This paper presents a centralized novel mobility prediction scheme which combines mobile node's location tracking through the analysis of RSSI samples and data mining through mobile nodes past movement history. The scheme works well with higher accuracy and low fluctuations with respect to frequency ranks. Typically, the movement history should be bigger for better prediction accuracy. Location tracking and data mining operates in parallel and is used when they are needed for prediction. The prediction accuracy of location tracking and data mining prediction scheme (LTDPS) looks quite satisfactory through our simulation analysis, as we compared its performance with other baseline schemes.

## ACKNOWLEDGMENT

This paper is an extended version of the paper submitted for CITA 2007 Conference, entitled –"A Novel Mobility Prediction in 802.11 Infrastructure Networks by Location Tracking and Data Mining", held during July 2007 in Kuching, Malaysia.